\documentclass[aps,superscriptaddress,twocolumn,twoside,floatfix,pra,a4paper]{revtex4-2}
\usepackage{times}
\usepackage{epsfig}
\usepackage{amsfonts}
\usepackage{amsmath}
\usepackage{amssymb,amsthm}
\usepackage{color}
\usepackage{multirow}
\usepackage{braket}
\usepackage{bbm}
\usepackage{latexsym}
\usepackage{amsfonts}
\usepackage{mathrsfs}
\usepackage{natbib}
\usepackage{verbatim}
\usepackage{gensymb}
\usepackage{caption}
\usepackage{subcaption}
\usepackage{subcaption}
\usepackage{graphicx}
\allowdisplaybreaks

\usepackage[colorlinks=true,linkcolor=blue,citecolor=magenta,urlcolor=blue]{hyperref}
\allowdisplaybreaks

\begin{document}

\title{How to Test Bell Nonlocality for Gravity?}

\author{Debarshi Das}
\email{dasdebarshi90@gmail.com}
\affiliation{Department of Physics, Shiv Nadar Institution of Eminence, Gautam Buddha Nagar, Uttar Pradesh 201314, India}

\author{Mir Alimuddin}
\affiliation{ICFO-Institut de Ciencies Fotoniques, The Barcelona Institute of Science and Technology, Av. Carl Friedrich Gauss 3, 08860 Castelldefels (Barcelona), Spain}

\author{Simon Storz}
\affiliation{Department of Physics, ETH Z\"{u}rich, 8093 Z\"{u}rich, Switzerland.}
\affiliation{Quantum Center, ETH Z\"{u}rich, 8093 Z\"{u}rich, Switzerland.}

\author{Yiwen Chu}
\affiliation{Department of Physics, ETH Z\"{u}rich, 8093 Z\"{u}rich, Switzerland.}
\affiliation{Quantum Center, ETH Z\"{u}rich, 8093 Z\"{u}rich, Switzerland.}

\author{Sougato Bose}
\affiliation{Department of Physics and Astronomy, University College London, Gower Street, London WC1E 6BT, England, United Kingdom}


\begin{abstract}
 We propose an experiment to test Bell nonlocality, a genuine nonclassicality, for the gravitational field. Two masses with embedded entangled spins (e.g., two diamonds with their NV-centre spins entangled) are placed well outside each other's light cones, ensuring a locality-loophole-free scenario. The spins are then coupled to the motion of their respective masses to generate spatial superpositions. Finally, local measurements are performed only on the gravitational fields of the two masses. If gravity is quantum, then the two entangled masses would entangle their gravitational fields, leading to correlations certifying Bell nonlocality of gravity. Trapped and ground-state cooled nano-objects with micron-sized spatial superposition are sufficient for this test. This goes beyond the recent proposals to test nonclassicality of gravity by providing, for the first time in the literature, a minimal tool to (i) create Einstein-Podolsky-Rosen (EPR) state of gravitational curvatures,    (ii) witness entangled gravitational curvatures, (iii) rule out any local-realist description of gravity, and (iv) achieve a loophole-free test of gravity's nonclassicality in a fully device-independent way.
\end{abstract}
\maketitle

{\em Introduction:}  To date, substantial progress has been made in developing both quantum and classical theories of gravitational field sourced by a quantum state of mass. However, empirically determining which of these theories describes our nature is still an open question. Laboratory experiments are ideal as they offer a greater degree of control and less chance of false signals from unknown sources, compared to, say, cosmological observations. Although one only has low-energy (infrared) probes in the laboratory, this can still be used to test the fundamental quantum nature of gravity. Computations of quantitative quantum corrections to gravitational interactions \cite{donoghue1995introduction}, or proposals to detect on-shell gravitons \cite{parikh2021signatures,parikh2021quantum,Tobar2024Detecting,Carney2024graviton} fall in this category, but remain challenging with near-term technologies. Recent advances in the quantum control of mesoscopic masses \cite{Yang2026mechanical,Omahen2025Ultracold,Piotrowski2023Simultaneous,Bild2023Schrodinger,Cattiaux2021Amacroscopic,Delic2020cooling,Rossi2018Measurement,Vovrosh2017parametric,OConnell2010Quantum} have led to proposals for low-energy tests of nonclassical features of gravity, such as its ability to mediate entanglement, and its other quantum signatures \cite{bose2016matter,bose2017spin,marletto2017gravitationally,qvarfort2020mesoscopic,krisnanda2022quantum,carney2021using,Biswas2022Gravitational,Etezad2024Paradox,Kent2022bell,Lami2024testing,hanif2024testing,Miki2025Role,Howl2021Non-Gaussianity,hanif2024testing,Strasser2025Evidencingquantum,bose2025spinbasedpathwaytestingquantum}. However, these approaches do not probe the strongest operational signature of quantumness--Bell nonlocality. Here we ask a sharper question: \textit{can gravitational curvatures themselves be prepared in a Bell-nonlocal, and hence, entangled state}? We propose a protocol, realizable in the foreseeable future, to test whether gravitational curvatures can give rise to Einstein-Podolsky-Rosen (EPR) correlations violating a Bell inequality, thereby ruling out all  local-realist (classical) descriptions of gravity in a loophole-free, fully device-independent way.

Bell nonlocality \cite{Bell1964on,Brunner2014Bell} represents the strongest form of spatial quantum correlations, as its existence implies the presence of other nonequivalent spatial quantum correlations, namely, EPR steering \cite{Uola2020quantum}, entanglement \cite{Horodecki2009quantum}, quantum discord \cite{Modi2014a,Bera2018quantum}, and quantum coherence \cite{Streltsov2017colloquium}. Consequently, the present proposal provides a means to test the existence of all possible spatial quantum correlations simultaneously within the context of gravity, bringing us closer to a more complete characterization of gravity as a quantum phenomenon.

The earlier-mentioned proposal based on quantum gravity-induced entanglement between masses (QGEM proposal) \cite{bose2016matter,bose2017spin,marletto2017gravitationally} aims to test whether gravitational interaction is capable of mediating quantum communication, assuming locality \cite{marshman2020locality} (or, no action at a distance -- not to be confused with the definition of locality in the context of Bell's inequality defined by EPR \cite{Einstein1935can}) of interaction. However, no further conclusions can be drawn from such a result. Interpretations such as the existence of quantum excitations of the gravitational vacuum (virtual gravitons) \cite{marshman2020locality}, operator-valued interactions \cite{bose2022mechanism}, and spin-$2$ nature of the gravitons \cite{Biswas2022Gravitational} provide the agent/mechanism for quantum communications via gravity. Other interpretations such as superposition of space-time curvatures \cite{christodoulou2019possibility,christodoulou2022locally} may offer sufficient explanations, but it is unclear whether they are also necessary to account for the observed entanglement between masses. Therefore, the QGEM cannot definitively conclude that it has detected entanglement {\em of} the gravitational field itself. 
After QGEM, while it would be a reasonable logical extrapolation to assume gravitational fields can also become entangled, that will not be equivalent to a loophole-free evidencing of entangled gravitational curvatures. Our proposal aims to bridge this gap by directly probing Bell nonlocality of the gravitational curvatures themselves, thereby providing operational evidence for entangled gravitational curvatures.


\begin{figure*}[ht]
    \centering
    \includegraphics[width=0.8\textwidth]{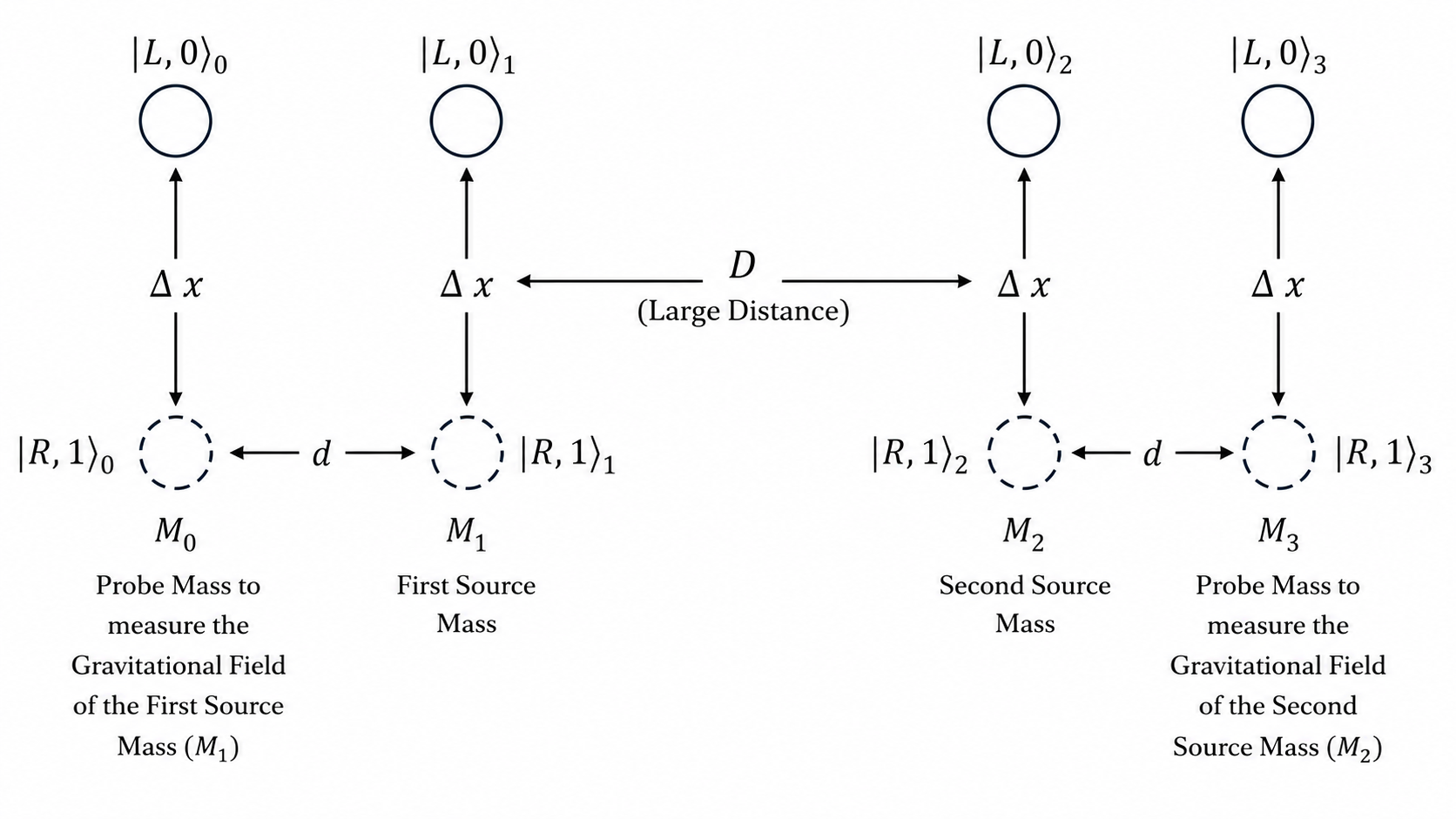}
    \caption{Geometry of the setup}
    \label{fig1}
\end{figure*}

If two test masses become entangled (like in QGEM \cite{bose2016matter,bose2017spin,marletto2017gravitationally}), it is conceivable that their associated gravitational curvatures may also become entangled. However, detecting such an entanglement would require performing local measurements on the gravitational curvatures themselves. This necessitates introducing two additional masses as probes, capable of interacting with and measuring local gravitational fields. Consequently, a total of at least four masses would be required: two entangled source masses to generate entanglement between their respective gravitational curvatures, and two probe masses that locally measure the corresponding gravitational fields to detect the entanglement. Our protocol explicitly incorporates these probe masses, providing the minimal architecture to enable Bell-test on the gravitational field itself. 



 Beyond detecting entangled gravitational curvatures, our protocol also differs fundamentally from the existing ones in terms of device-independence. For example, the QGEM is not a device-independent test of the nonclassicality of gravity. Further, the correlations generated in this experiment can be reproduced by local-realist models, as they do not involve measurements on spacelike-separated masses. On the other hand, the recent proposal \cite{hanif2024testing} avoids the need for trusted measurement devices, but eliminating loopholes (particularly, the classical disturbance loophole) still hinges on technological advances. In contrast, the present proposal is a fully device-independent test of the nonclassicality of gravity. The locality loophole is closed by construction, since the Bell test is performed on gravitational fields sourced by masses that are spacelike separated (i.e., outside one another's light cones). 

\textit{Details of the proposal:} Now, the details of the proposals will be presented (for Mathematical details, see Appendix \ref{A1}). Let us consider that there are two masses $M_1$ and $M_2$ (each with mass $M$) which are located far away from each other (the distance between the masses will be estimated later). Each of these masses has embedded spins (spin-$1/2$). The spin degrees of freedom of these two masses are denoted by $S_1$ and $S_2$ respectively. In the following, the steps of the proposal will be explained.

\textbf{Step 1: Entangling Distant Spins--} The two masses initially are in localized states $|\xi\rangle_{M_1}$ and $|\xi\rangle_{M_2}$ respectively.  The two spins embedded in the two masses are entangled by some means (details will be presented later, under physical realisations). Hence, the initial state of the whole system becomes
\begin{align}
    |\psi_1\rangle_{S_1 S_2} = \frac{1}{\sqrt{2}}\left(|0 \rangle_{S_1} |0 \rangle_{S_2} + |1 \rangle_{S_1} |1 \rangle_{S_2}\right) \otimes |\xi\rangle_{M_1} \otimes |\xi\rangle_{M_2}, \nonumber
\end{align}
where $|0\rangle$ and $|1\rangle$ are the eigenstates of $\sigma_z$. 

\textbf{Step 2: Entangling Spatial Degrees of Freedom--} Each of these masses is then prepared in spatial superposition via the following unitary evolution (by subjecting each spin to an inhomogeneous magnetic field \cite{Bose2025massive,bose1999scheme,scala2013matter,wan2016free,pedernales2020motional,marshman2022constructing,margalit2021realization,zhou2022catapulting,Zhou2023mass}):
\begin{align}
|\xi\rangle_{M_i} \otimes |0 \rangle_{S_i} \rightarrow |L, 0 \rangle_i, \hspace{0.5cm} |\xi\rangle_{M_i} \otimes |1 \rangle_{S_i} \rightarrow |R, 1 \rangle_i 
\label{unitary}
\end{align}
with $i=1,2$. In the above, the states $|L, 0 \rangle_i$ and $|R, 1 \rangle_i$ for each $i \in \{1,2\}$ are separated by a distance $\Delta x$. After attaining these superpositions, the inhomogeneous magnetic fields are switched off.

Hence, the joint state of the two masses becomes
\begin{align}
    |\psi_2\rangle_{1 2} = \frac{1}{\sqrt{2}}\left(|L,0 \rangle_{1} |L, 0 \rangle_{2} + |R, 1 \rangle_{1} |R, 1 \rangle_{2}\right).
    \label{entangledsource}
\end{align}
Since the distance between the two source masses $M_1$ and $M_2$ is very large, all interactions between them are negligible. If gravitational interaction is quantum (i.e., it is operator-valued at each point, instead of number-valued), the state in Eq.(\ref{entangledsource}) above is equivalent to entangled geometries (curvatures) -- this is the resource of our Bell experiment. 

\textbf{Step 3: Probe Preparation--} Two more masses $M_0$ and $M_3$ (each with mass $m$) with embedded spins are now considered. The mass $M_0$ ($M_2$) is located near $M_1$ ($M_3$) which will be used as a probe to measure the gravitational field of $M_1$ ($M_3$). The spin degrees of freedom of $M_0$ and $M_3$ are denoted by $S_0$ and $S_3$ respectively. These two masses and their spins are uncorrelated initially.

Each of the spins $S_0$ and $S_3$ is initially in a superposed state:
\begin{align}
|\phi\rangle_{S_i} = \frac{1}{\sqrt{2}}\left(|0\rangle_{s_i} +  |1\rangle_{s_i}\right) \, \, \, \, \text{with} \, \, i = 0,3. \nonumber
\end{align}

Each of these masses $M_0$ and $M_3$ is then prepared in spatial superposition via the above-mentioned unitary evolution (\ref{unitary}) with $i=0,3$.
Here also, the states $|L, 0 \rangle_i$ and $|R, 1 \rangle_i$ for each $i \in \{0,3\}$ are separated by a distance $\Delta x$. Although we have assumed the same superposition size $\Delta x$ for all masses for simplifying the analytical calculations, this is not a strict requirement of our proposal.

The separation between the state  $|L, 0 \rangle_0$ ($|R, 1 \rangle_0$) of $M_0$ and the state $|L, 0 \rangle_1$ ($|R, 1 \rangle_1$) of $M_1$ is $d$. Similarly,  the separation between the state  $|L, 0 \rangle_2$ ($|R, 1 \rangle_2$) of $M_2$ and the state $|L, 0 \rangle_3$ ($|R, 1 \rangle_3$) of $M_3$ is $d$. The geometry of the setup is presented in Fig. \ref{fig1}.

\textbf{Step 4: Source-Probe Gravitational Interaction--} Now, the masses $M_0$ and $M_1$ interact through gravity only for a time duration $\tau$. Similarly, the masses $M_2$ and $M_3$ interact only through gravity for the same duration. The gravitational interactions between the other pairs of masses are negligible. Although challenging, electromagnetic interaction between $M_0$ and $M_1$ or between $M_2$ and $M_3$ can, in principle, be made negligible by neutralizing them and/or placing electromagnetic screening between them \cite{van2020quantum,schut2023micron,schut2022deco,Elahi2025Diamagnetic}.


$\bullet$ \textbf{Step 5: Measuring the spins $S_1$ and $S_2$ (local operations)--} The spatial superpositions of the masses $M_1$ and $M_2$ are closed by reversing the local unitary dynamics (\ref{unitary}) with $i=1,2$. This can be achieved with a magnetic field homogeneity oriented oppositely to the apparatus in step 2, or simply by flipping
the spin. Each of the spins $S_1$ and $S_2$ are measured in the basis $|\pm\rangle_{S_i} = (|0\rangle_{S_i} \pm |1\rangle_{S_i})/\sqrt{2}$ with $i=1,2$. 

$\bullet$ \textbf{Step 6: Measuring the spins $S_0$ and $S_3$ (measurements for Bell test)--} The spatial superpositions of the masses $M_0$ and $M_3$ are now closed by reversing the local unitary dynamics (\ref{unitary}) with $i=0,3$. Next, local spin measurements corresponding the operators $A_0 = \sigma_x$ and $A_1 = -\sigma_z$  are performed on $S_0$, and local spin measurements corresponding the operators $B_0 = (\sigma_x+\sigma_z)/\sqrt{2}$ and $B_1 = (\sigma_x-\sigma_z)/\sqrt{2}$ are performed on $S_3$. 
The statistics thus collected are used to test Bell nonlocality. 

The Bell-Clauser–Horne–Shimony–Holt (Bell-CHSH) inequality \cite{Clauser1969proposed} is given by,
\begin{equation}
    B = \langle A_0 B_0 \rangle + \langle A_0 B_1 \rangle + \langle A_1 B_0 \rangle - \langle A_1 B_1 \rangle \leq 2.
\end{equation}
Violation of this inequality implies Bell nonlocality -- no local-realist description of the system (or, field) on which the local measurements are performed.

Now, the following results are derived considering that gravity is quantum in nature (in the low energy limit, this is equivalent to considering the Newtonian potential in operator form \cite{Carney2019tabletop}). When the results $++$ or $--$ are obtained in Step 5 by measuring on the spins $S_1$ and $S_2$, then it can be shown that 
\begin{align}
 |\langle \bar{B} \rangle|_{++} = |\langle \bar{B} \rangle|_{--} = &\left|\frac{4 \sqrt{2}}{\cos (2 \theta )+3}\right|. 
 \label{bell1}
\end{align}
Here 
\begin{align}
    \theta = \frac{G M m \tau}{\hbar \sqrt{d^2 + \Delta x^2}} - \frac{G M m \tau}{\hbar d}.
    \label{theta}
\end{align}
These cases occur with probability $P(++) = P(--) =   (1 + \cos^2 \theta)/4 \neq 0$ for any $\theta$. On the other hand, when the outcomes $+-$ or $-+$ are obtained in Step 5, then
\begin{align}
 |\langle \bar{B} \rangle|_{+-} = |\langle \bar{B} \rangle|_{-+} = 2 \sqrt{2} > 2 \, \, \text{for any} \, \, \theta.   
 \label{bell2}
\end{align}
These cases occur with probabilities $P(+-) = P(-+) = \sin^2\theta/4 \neq 0$ for any $\theta \neq n \pi$ ($n$ being arbitrary integer).

{\centering
	\begin{table*}[ht!]
		\begin{tabular}{ |c|c|c|c|c|c|c|c|c| } 
			\hline
			$M$ (kg) & $m$ (kg) & $d$ (m) & $\Delta x$ (m) & $\tau$ (s) & $|\langle \bar{B} \rangle|_{++} = |\langle \bar{B} \rangle|_{--}$ &  $P(++) = P(--)$ & $|\langle \bar{B} \rangle|_{+-} = |\langle \bar{B} \rangle|_{-+}$ &  $P(+-) = P(-+)$ \\
			\hline
            \hline
			$5 \times 10^{-14}$ & $2 \times 10^{-14}$  & $10^{-4}$ & $10^{-4}$ & $1$ & $2.62>2$ & $0.27$ & $2 \sqrt{2} > 2$ & $0.23$  \\
            \hline
            $10^{-13}$ & $2 \times 10^{-14}$  & $2 \times 10^{-4}$ & $2 \times 10^{-4}$ & $1$ & $2.62>2$ & $0.27$ & $2 \sqrt{2} > 2$ & $0.23$  \\
            \hline
            $10^{-13}$ & $2 \times 10^{-14}$  & $10^{-4}$ & $10^{-4}$ & $0.5$ & $2.62>2$ & $0.27$ & $2 \sqrt{2} > 2$ & $0.23$  \\
			\hline
            $10^{-14}$ & $10^{-14}$  & $10^{-5}$ & $10^{-5}$ & $1$ & $2.62>2$ & $0.27$ & $2 \sqrt{2} > 2$ & $0.23$  \\
            \hline
            $10^{-14}$ & $10^{-14}$  & $10^{-6}$ & $10^{-6}$ & $0.1$ & $2.62>2$ & $0.27$ & $2 \sqrt{2} > 2$ & $0.23$  \\
			\hline
            $10^{-13}$ & $1.8 \times 10^{-13}$  & $2 \times 10^{-4}$ & $2 \times 10^{-4}$ & $1$ & $2.15>2$ & $0.33$ & $2 \sqrt{2} > 2$ & $0.17$  \\
			\hline
            $10^{-13}$ & $1.2 \times 10^{-13}$  & $2 \times 10^{-4}$ & $2 \times 10^{-4}$ & $1$ & $2.78>2$ & $0.25$ & $2 \sqrt{2} > 2$ & $0.25$  \\
			\hline
		\end{tabular}
		\caption{Representative parameter sets for the proposed protocol. Row 1 shows the reference parameters. Row 2 uses a larger source mass and larger separation. Row 3 compensates for a shorter interaction time by increasing the source mass. Rows 4 and 5 demonstrate the feasibility with smaller masses and correspondingly smaller separations (and a very short interaction time in Row 6). Rows 6 and 7 illustrate the effect of varying the probe-to-source mass ratio. Note that experimental demonstration of either $|\langle \bar{B} \rangle|_{++} = |\langle \bar{B} \rangle|_{--}$ or $|\langle \bar{B} \rangle|_{+-} = |\langle \bar{B} \rangle|_{-+}$ is sufficient for our purpose. } \label{tab1}
	\end{table*}
}

The Step 4 along with the Step 6 mentioned earlier constitute measurements on the gravitational field of $M_1$ and $M_2$ by probe masses $M_0$ and $M_3$ respectively \cite{hanif2024testing}. Note that any measurement consists of two parts-- (1) the `measurement interaction part', which is responsible for transferring information from the system to be measured to the probe, and (2) the `reading out' part, which gives access of the above information by reading the outcome of the measurement. In the present context, Step 4 is the `measurement interaction part' where each probe interacts with its source only via gravity, whereas Step 6 is the `reading out' part. 
After closing the superposition of each probe, the probe spin decouples from its spatial degrees of freedom, carrying the information about the relative phase acquired between $|L,0\rangle_i$ and $|R,1\rangle_i$ ($i=1,2$) due to the mutual gravitational interaction between the source and the probe. A projective measurement of the probe spin thus reveals the information about the aforementioned relative phases (and, consequently, about the mutual gravitational interactions). Hence, the spin projective measurement realizes a POVM on the gravitational field of the source mass. Similar justification in quantum field theory language is given in \cite{Polo-Gomez2022detector}.

Step 5 involves only local measurements. Hence, it cannot create any additional entanglement accross the bipartition $M_0-M_1$ vs. $M_2-M_3$. 
Thus the measurement used for post-selection in Step 5 does not itself generate the Bell violation out of local models between two distant labs (see Appendix \ref{A2})--it merely reveals the Bell nonlocality of the gravitational fields in the post-selected branches. Moreover, the measurement settings in Step 6 are chosen independently of the outcomes obtained in Step 5. 
Crucially, the local spin measurements on the source masses in Step 5 are performed only after the ‘measurement interaction’ stage (Step 4), i.e.,  after each system (source mass's gravitational field) has already been correlated with the corresponding probe.

In this proposal, the quantum correlations of the entangled source masses is imprinted onto their respective gravitational fields, which in turn mediate correlations to the probe masses. Local measurements on the sources then play an assisting role: they herald post-selected branches in which the gravitational fields yield Bell-nonlocal correlations. In this way, the process realizes an assisted form of Bell nonlocality of the gravitational fields, with the probes serving as readouts of the quantum correlations carried by the fields of entangled masses. 

In particular, the measurements performed in Step 5 are fixed local measurements, and their outcomes are not used to choose the Bell-test measurement settings in Step 6. In Appendix \ref{A2}, we show that this action of `fair' post-selection cannot generate Bell nonlocality, out of local models. 

\textit{Parameter Estimation--} For \(d\gg\Delta x\), we have 
\(\theta\rightarrow 0\), and the resulting correlations are insufficient to 
produce a Bell violation. On the other hand, the regime \(d\ll\Delta x\) 
is experimentally 
challenging due to the difficulty of maintaining coherent control while 
suppressing short-range interactions simultaneously. We therefore consider the intermediate 
regime \(d\approx\Delta x\). In this regime, small values of \(\theta\) (i.e., smaller $M$ and $m$) lead to negligible Bell 
violation (see Eqs.(\ref{bell1})-(\ref{bell2})), and hence the objective is to engineer the experimental parameters 
such that the phase approaches the optimal values , 
\(\theta\simeq\pi/2\) (mod $2 \pi$), where the Bell signal is maximized.

In Table \ref{tab1} some representative parameter values suitable for this experiment is presented. This proposal is implementable with typical masses $\sim 10^{-13}-10^{-14}$ kg having spatial superposition of sizes $\Delta x \sim 1-100 \mu$m. Increasing further the masses makes preparing the superposition prohibitive.


\textit{Physical Realization--} Micro-diamonds with an
embedded NV centre spin  is one possible candidate system for each mass (typical mass $\sim 10^{-13}-10^{-15}$ kg) with embedded spin. Such micro-diamonds can be trapped in various low-noise traps based on Levitated mechanics \cite{Gonzalez-Ballestero2021Levitodynamics} and can be effectively cooled to ground state through feed-back cooling and other techniques \cite{Gieseler2012Subkelvin,Doherty1999Feedback,Walker2019Measurement,Vinante2019Testing,Tebbenjohanns2021Quantum,Magrini2021Real-time}. Another interesting platform might be magnetically levitated microspheres \cite{Romero-Isart2012Quantum,Timberlake2021Probing,Hansen2026Optical}. 

The measurement times ($\sim \tau =$ interaction time between each pair of the source and the probe masses) for the two distant masses $M_1$ and $M_2$ with embedded spins are typically $\sim 1\,\mathrm{s}$ and $\sim 0.1\,\mathrm{s}$. To strictly close the locality loophole, the entangled spins of the two masses $M_1$ and $M_2$ must be separated by a distance such that no signal can propagate between them during the measurement time $\sim \tau$. Since the speed of light is approximately $3\times 10^5$ km/s, 
the required separation between the spins should be $\sim 10^4-10^5$ km. Satellite-based experiments have already demonstrated distribution of entangled photon pair  over distances exceeding $1200$ km \cite{Yin2017Satellite}. Furthermore, theoretical proposals show that such entangled photon pairs can be distributed over distances $\sim 10^5$ km in the foreseeable future \cite{Rideout2018Bell}. These results establish photons as reliable carriers of entanglement across astronomical distances-- the key requirement in our proposal. Further, for such a large distance, the freedom-of-choice loophole can also be closed, even using human free choice \cite{Rideout2018Bell}.  

The next step is to transfer this photonic entanglement into the spins of NV centres in diamonds. This can be achieved by coupling the photons with solid-state spin-qubits, as demonstrated in NV centres \cite{Togan2010Quantum}. More generally, protocols for mapping photonic entanglement into matter systems have been realized in atomic and solid-state systems \cite{Kimble2008The,Simon2008Quantum,Clausen2011Quantum,Tiranov2015Rare,Sangouard2013Heralded}. These works show that photonic entanglement can be converted into spin entanglement between two distant masses. Therefore, by combining satellite-based photonic entanglement distribution  with efficient spin--photon coupling, it is expected to realize entanglement between spins in distant NV centre diamonds. Alternatively, two distant spins of NV centres can be entangled in a heralded manner by joint detection of indistinguishable photons from the NV-centre sources \cite{Bernien2013Heralded,Bose1999Proposal}.

After establishing the entanglement between the spins of the two source masses, steps 2-6 should be followed. To ensure that steps 4 and 6 constitute the measurements on the gravitational field of the source masses $M_1$ and $M_2$, the distance between each pair of source-probe ($M_0-M_1$ and $M_2-M_3$) is such that the gravitational interaction between each pair is much stronger than the electromagnetic interactions. For completely neutral diamond nano-crystals, this can be achieved for $d > 157 \mu$m  \cite{van2020quantum}. For smaller $d$, electromagnetic screening between each source-probe pair will be required \cite{van2020quantum,schut2023micron,schut2022deco,Elahi2025Diamagnetic}.

Note that detecting entangled gravitational curvature alone, without resorting to any Bell test, can be achieved in a similar setup, where Step 6 mentioned earlier is replaced by performing an entanglement witness test through measurements of the spins $S_0$ and $S_3$. Since the locality loophole is not relevant for entanglement witness tests, the two source masses need not be placed outside each other's light cones (i.e., within a single laboratory is sufficient), which significantly eases the experimental demands. This alternative has experimental requirements comparable to those of QGEM, while providing the additional evidence of entangled space-time geometries.

\textit{Conclusions--} Testing whether gravity is a quantum entity remains one of the biggest challenges of modern science. At the same time, loophole-free demonstrations of Bell nonlocality stand among the greatest achievements in quantum science and technology. In this work, we have proposed an experiment that brings together these two parallel yet largely independent research directions, aiming to test whether gravity exhibits quantumness in the form of Bell nonlocality. Such a proposal appears realisable in the future, given ongoing experimental advances in long-distance distribution of entangled pair of photons, photon–spin interfaces, and the trapping and ground-state cooling of mesoscopic masses. The scope of this proposal extends beyond testing the quantumness of gravity: it aims to demonstrate Bell nonlocality and entanglement between gravitational curvatures--arguably among the most ``spooky'' quantum features in the gravitational context.

Before concluding, it is to be noted that the recent proposal by Kent \textit{et al.} \cite{Kent2022bell}--an extension of the QGEM proposal--aims to detect the gravity-induced entanglement between two masses through a Bell test (i.e., in a device-independent way). Hence, this proposal again aims to test the ability of gravitational interaction to act as a quantum communication channel, but in a more loophole-free way. Specifically, this proposal cannot test Bell nonlocality (and, entanglement) of the gravitational curvatures themselves.

\begin{acknowledgments}
{\it Acknowledgements.--}
DD and SB acknowledge the financial support from the Royal Society, UK under the scheme ``Newton International Fellowships
Alumni 2025'' (Grant No. AL$\backslash$251043). DD gratefully acknowledges University College London for its kind hospitality during his visit in July 2026, which provided the opportunity to complete a substantial part of this work. M.A. acknowledges funding from the European Union (QURES, 101153001). S.S. and Y.C. acknowledge funding from the National Research Council of Science \& Technology (NST) grant by the Korea government (MSIT) (No. GTL25011-000). SB would like to
acknowledge EPSRC grant EP/X009467/1 and STFC grant
ST/W006227/1. This work was made possible through
the support of the WOST, WithOut SpaceTime project
(https://withoutspacetime.org), supported by Grant ID\#63683
from the John Templeton Foundation (JTF). S.B’s research is
funded by the Gordon and Betty Moore Foundation through
Grant GBMF12328, DOI 10.37807/GBMF12328, and the Alfred P. Sloan Foundation under Grant No. G-2023-21130.
\end{acknowledgments}

\bibliography{ref2} 

\appendix

\onecolumngrid

\section{Mathematical Details:} \label{A1}
Let us consider that there are two masses $M_1$ and $M_2$ (each with mass $M$) which are located far away from each other. Each of these masses has embedded spins (spin-$1/2$). The spin degrees of freedom of these two masses are denoted by $S_1$ and $S_2$ respectively. 

$\bullet$ \textbf{Step 1} Entangling Distant Spins: These two spins get entangled by some means. Let the initial entangled state of the two spins is given by,
\begin{align}
    |\psi_1\rangle_{S_1 S_2} = \frac{1}{\sqrt{2}}\left(|0 \rangle_{S_1} |0 \rangle_{S_2} + |1 \rangle_{S_1} |1 \rangle_{S_2}\right),
\end{align}
where $|0\rangle$ and $|1\rangle$ are the eigenstates of $\sigma_z$. Since each of the two masses initially is in localized states $|\xi\rangle_{M_1}$ and $|\xi\rangle_{M_2}$, the initial joint state of the spatial and spin degrees of freedom of $M_1$ and $M_2$ is given by,
\begin{align}
    |\psi_1\rangle_{1 2} = |\xi\rangle_{M_1} \otimes \frac{1}{\sqrt{2}}\left(|0 \rangle_{S_1} |0 \rangle_{S_2} + |1 \rangle_{S_1} |1 \rangle_{S_2}\right) \otimes |\xi\rangle_{M_2}.
\end{align}

$\bullet$ \textbf{Step 2} Entangling Spatial Degrees of Feedoms: Each of these spins is then prepared in spatial superposition via the unitary evolution (by subjecting each spin to an inhomogeneous magnetic field):
\begin{align}
|\xi\rangle_{M_i} \otimes |0 \rangle_{S_i} \rightarrow |L, 0 \rangle_i, \hspace{0.5cm} |\xi\rangle_{M_i} \otimes |1 \rangle_{S_i} \rightarrow |R, 1 \rangle_i  \, \, \, \, \text{with} \, \, i = 1,2.
\end{align}
In the above, the states $|L, 0 \rangle_i$ and $|R, 1 \rangle_i$ for each $i \in \{1,2\}$ are separated by a distance $\Delta x$. After attaining these superpositions, the inhomogeneous magnetic fields are switched off.

Hence, the joint state of the two masses becomes
\begin{align}
    |\psi_2\rangle_{1 2} = \frac{1}{\sqrt{2}}\left(|L,0 \rangle_{1} |L, 0 \rangle_{2} + |R, 1 \rangle_{1} |R, 1 \rangle_{2}\right).
\end{align}

Here we have assumed that no interaction is present between the two masses. Since the distance between the two masses is very large, all interactions between them are negligible. 

$\bullet$ \textbf{Step 3}: Probe Preparation: Two more masses $M_0$ and $M_3$ (each with mass $m$) with embedded spins are now considered.  The spin degrees of freedom of $M_0$ and $M_3$ are denoted by $S_0$ and $S_3$ respectively. These two masses and their spins are uncorrelated initially.

Each of the spins $S_0$ and $S_3$ is initially in a superposed state:
\begin{align}
|\phi\rangle_{S_i} = \frac{1}{\sqrt{2}}\left(|0\rangle_{s_i} +  |1\rangle_{s_i}\right) \, \, \, \, \text{with} \, \, i = 0,3.
\end{align}

Each of these spins $S_0$ and $S_3$ is prepared in spatial superposition via the unitary evolution (by subjecting each spin to an inhomogeneous magnetic field):
\begin{align}
|\xi\rangle_{M_i} \otimes |0 \rangle_{S_i} \rightarrow |L, 0 \rangle_i, \hspace{0.5cm} |\xi\rangle_{M_i} \otimes |1 \rangle_{S_i} \rightarrow |R, 1 \rangle_i  \, \, \, \, \text{with} \, \, i = 0,3.
\end{align}
Here, the states $|L, 0 \rangle_i$ and $|R, 1 \rangle_i$ for each $i \in \{0,3\}$ are separated by a distance $\Delta x$. Again, after creating these superpositions, the inhomogeneous magnetic fields are switched off.

Now, the mass $M_0$ is brought near $M_1$ which will be used as a probe to measure the gravitational field of $M_1$. Similarly, $M_3$ is brought near $M_2$ for using as a probe to measure the gravitational field of $M_2$. 

The separation between the state  $|L, 0 \rangle_0$ ($|R, 1 \rangle_0$) of $M_0$ and the state $|L, 0 \rangle_1$ ($|R, 1 \rangle_1$) of $M_1$ is $d$. Similarly,  the separation between the state  $|L, 0 \rangle_2$ ($|R, 1 \rangle_2$) of $M_2$ and the state $|L, 0 \rangle_3$ ($|R, 1 \rangle_3$) of $M_3$ is $d$. The geometry of the setup is presented in Fig. \ref{fig1}.

At this stage, the joint state of the four masses becomes
\begin{align}
      |\psi_3\rangle_{0 1 2 3} = \frac{1}{\sqrt{2}}\left(|L,0\rangle_{0} +  |R, 1\rangle_{0}\right) \otimes \frac{1}{\sqrt{2}}\left(|L,0 \rangle_{1} |L, 0 \rangle_{2} + |R, 1 \rangle_{1} |R, 1 \rangle_{2}\right) \otimes \frac{1}{\sqrt{2}}\left(|L,0\rangle_{3} +  |R, 1\rangle_{3}\right).
\end{align}

$\bullet$ \textbf{Step 4}: Source-Probe Gravitational Interaction: Now, the masses $M_0$ and $M_1$ interact through gravity only. Similarly, the masses $M_2$ and $M_3$ interact only through gravity. The gravitational interactions between the other pairs of masses are negligible. Electromagnetic interaction between $M_0$ and $M_1$ or between $M_2$ and $M_3$ can be made negligible by neutralizing them and/or placing electromagnetic screening between them.

Let us assume that the pairs $M_0$-$M_1$ and $M_2$-$M_3$ interact gravitationally for a time duration $\tau$. If gravity is quantum in nature (in the low energy limit, this is equivalent to considering the Newtonian potential in operator form \cite{Carney2019tabletop}) the joint state of the four masses becomes (neglecting the global phase $e^{2i \theta_2}$)
\begin{align}
      |\psi_4\rangle_{0 1 2 3} = \frac{1}{2 \sqrt{2}}\Big(&|L,0\rangle_{0} |L,0\rangle_{1} |L,0\rangle_{2} |L,0\rangle_{3} + e^{2i \theta} |L,0\rangle_{0} |R,1\rangle_{1} |R,1\rangle_{2} |L,0\rangle_{3} \nonumber \\
      &+ e^{i \theta} |R,1\rangle_{0} |L,0\rangle_{1} |L,0\rangle_{2} |L,0\rangle_{3} + e^{i \theta} |R,1\rangle_{0} |R,1\rangle_{1} |R,1\rangle_{2} |L,0\rangle_{3} \nonumber \\
      & + e^{i \theta} |L,0\rangle_{0} |L,0\rangle_{1} |L,0\rangle_{2} |R,1\rangle_{3} + e^{i \theta} |L,0\rangle_{0} |R,1\rangle_{1} |R,1\rangle_{2} |R,1\rangle_{3}\nonumber \\
      & + e^{2 i \theta} |R,1\rangle_{0} |L,0\rangle_{1} |L,0\rangle_{2} |R,1\rangle_{3} +  |R,1\rangle_{0} |R,1\rangle_{1} |R,1\rangle_{2} |R,1\rangle_{3}\Big),
      \label{statestep4}
\end{align}
where 
\begin{align}
    \theta = \theta _{1} -\theta_2 \nonumber 
\end{align}
with
\begin{align}
    \theta_1 &= \frac{G M m \tau}{\hbar \sqrt{d^2 + \Delta x^2}}, \nonumber \\
    \theta_2 &= \frac{G M m \tau}{\hbar d}.
\end{align}

$\bullet$ \textbf{Step 5: Measuring the spins S1 and S2 (local operations)--} The spatial superpositions of the masses $M_1$ and $M_2$ are closed following the below local unitary evolutions:
\begin{align}
 |L, 0 \rangle_i \rightarrow |\xi\rangle_{M_i} \otimes |0 \rangle_{S_i}, \hspace{0.5cm}  |R, 1 \rangle_i \rightarrow |\xi\rangle_{M_i} \otimes |1 \rangle_{S_i}  \, \, \, \, \text{with} \, \, i = 1,2.
\end{align}
Hence, the state (\ref{statestep4}) becomes:
\begin{align}
      |\psi_5\rangle_{0 1 2 3} = \frac{1}{2 \sqrt{2}}\Big(&|L,0\rangle_{0} |0\rangle_{S_1} |0\rangle_{S_2} |L,0\rangle_{3} + e^{2i \theta} |L,0\rangle_{0} |1\rangle_{S_1} |1\rangle_{S_2} |L,0\rangle_{3} \nonumber \\
      &+ e^{i \theta} |R,1\rangle_{0} |0\rangle_{S_1} |0\rangle_{S_2} |L,0\rangle_{3} + e^{i \theta} |R,1\rangle_{0} |1\rangle_{S_1} |1\rangle_{S_2} |L,0\rangle_{3} \nonumber \\
      & + e^{i \theta} |L,0\rangle_{0} |0\rangle_{S_1} |0\rangle_{S_2} |R,1\rangle_{3} + e^{i \theta} |L,0\rangle_{0} |1\rangle_{S_1} |1\rangle_{S_2} |R,1\rangle_{3}\nonumber \\
      & + e^{2 i \theta} |R,1\rangle_{0} |0\rangle_{S_1} |0\rangle_{S_2} |R,1\rangle_{3} +  |R,1\rangle_{0} |1\rangle_{S_1} |1\rangle_{S_2} |R,1\rangle_{3}\Big) \otimes |\xi\rangle_{M_1} \otimes |\xi\rangle_{M_2},
      \label{statestep5}
\end{align}

Each of the spins of the masses $M_1$ and $M_2$ are measured in the basis $|\pm\rangle_{S_i} = (|0\rangle_{S_i} \pm |1\rangle_{S_i})/\sqrt{2}$ with $i=1,2$. The post-measurement states are given by,
\begin{align}
    |\psi^{++}_6\rangle_{1,2,3,4} =& \frac{1}{\sqrt{3 + \cos 2 \theta}} \Big( \cos \theta |L,0\rangle_{0} |L,0\rangle_{3} + |R,1\rangle_{0} |L,0\rangle_{3} + |L,0\rangle_{0} |R,1\rangle_{3} + \cos \theta |R,1\rangle_{0} |R,1\rangle_{3} \Big) \nonumber \\
    & \hspace{2.5cm} \otimes \frac{|0\rangle_{S_1} + |1\rangle_{S_1}}{\sqrt{2}} \otimes |\xi\rangle_{M_1} \otimes \frac{|0\rangle_{S_2} + |1\rangle_{S_2}}{\sqrt{2}} \otimes |\xi\rangle_{M_2},
    \label{p1}
\end{align}
\begin{align}
    |\psi^{+-}_6\rangle_{1,2,3,4} =& \frac{1}{\sqrt{2}} \Big( |L,0\rangle_{0} |L,0\rangle_{3} -  |R,1\rangle_{0} |R,1\rangle_{3} \Big) \otimes \frac{|0\rangle_{S_1} + |1\rangle_{S_1}}{\sqrt{2}} \otimes |\xi\rangle_{M_1} \otimes \frac{|0\rangle_{S_2} - |1\rangle_{S_2}}{\sqrt{2}} \otimes |\xi\rangle_{M_2},
    \label{p2}
\end{align}
\begin{align}
    |\psi^{-+}_6\rangle_{1,2,3,4} =& \frac{1}{\sqrt{2}} \Big( |L,0\rangle_{0} |L,0\rangle_{3} -  |R,1\rangle_{0} |R,1\rangle_{3} \Big) \otimes \frac{|0\rangle_{S_1} - |1\rangle_{S_1}}{\sqrt{2}} \otimes |\xi\rangle_{M_1} \otimes \frac{|0\rangle_{S_2} + |1\rangle_{S_2}}{\sqrt{2}} \otimes |\xi\rangle_{M_2},
    \label{p3}
\end{align}
\begin{align}
    |\psi^{--}_6\rangle_{1,2,3,4} =& \frac{1}{\sqrt{3 + \cos 2 \theta}} \Big( \cos \theta |L,0\rangle_{0} |L,0\rangle_{3} + |R,1\rangle_{0} |L,0\rangle_{3} + |L,0\rangle_{0} |R,1\rangle_{3} + \cos \theta |R,1\rangle_{0} |R,1\rangle_{3} \Big) \nonumber \\
    & \hspace{2.5cm} \otimes \frac{|0\rangle_{S_1} - |1\rangle_{S_1}}{\sqrt{2}} \otimes |\xi\rangle_{M_1} \otimes \frac{|0\rangle_{S_2} -|1\rangle_{S_2}}{\sqrt{2}} \otimes |\xi\rangle_{M_2},
    \label{p4}
\end{align}
where $|\psi^{ij}_6\rangle_{1,2,3,4}$ is the post-measurement state, when the outcomes $i$ and $j$ are obtained for spins $S_1$ and $S_2$ respectively.

The outcome probabilities are given by,
\begin{align}
    &P(++) = P(--) = \frac{1 + \cos^2 \theta}{4}, \nonumber \\
    &P(+-) = P(-+) = \frac{\sin^2\theta}{4}.
\end{align}

$\bullet$ \textbf{Step 6: Measuring the spins $S_0$ and $S_3$ (measurements for Bell test)--} The spatial superpositions of the masses $M_0$ and $M_3$ are now closed following the below local unitary evolutions:
\begin{align}
 |L, 0 \rangle_i \rightarrow |\xi\rangle_{M_i} \otimes |0 \rangle_{S_i}, \hspace{0.5cm}  |R, 1 \rangle_i \rightarrow |\xi\rangle_{M_i} \otimes |1 \rangle_{S_i}  \, \, \, \, \text{with} \, \, i = 0,3.
\end{align}
Hence, the states (\ref{p1})-(\ref{p4}) become:
\begin{align}
    |\psi^{++}_7\rangle_{1,2,3,4}  =& \frac{1}{\sqrt{3 + \cos 2 \theta}} \Big( \cos \theta |0\rangle_{S_0} |0\rangle_{S_3} + |1\rangle_{S_0} |0\rangle_{S_3} + |0\rangle_{S_0} |1\rangle_{S_3} + \cos \theta |1\rangle_{S_0} |1\rangle_{S_3} \Big) \nonumber \\
    & \hspace{0.2cm} \otimes |\xi\rangle_{M_0} \otimes |\xi\rangle_{M_3} \otimes \frac{|0\rangle_{S_1} + |1\rangle_{S_1}}{\sqrt{2}} \otimes |\xi\rangle_{M_1} \otimes \frac{|0\rangle_{S_2} + |1\rangle_{S_2}}{\sqrt{2}} \otimes |\xi\rangle_{M_2},
    \label{p12}
\end{align}
\begin{align}
     |\psi^{--}_7\rangle_{1,2,3,4} =& \frac{1}{\sqrt{3 + \cos 2 \theta}} \Big( \cos \theta |0\rangle_{S_0} |0\rangle_{S_3} + |1\rangle_{S_0} |0\rangle_{S_3} + |0\rangle_{S_0} |1\rangle_{S_3} + \cos \theta |1\rangle_{S_0} |1\rangle_{S_3} \Big) \nonumber \\
    & \hspace{0.2cm} \otimes |\xi\rangle_{M_0} \otimes |\xi\rangle_{M_3} \otimes \frac{|0\rangle_{S_1} - |1\rangle_{S_1}}{\sqrt{2}} \otimes |\xi\rangle_{M_1} \otimes \frac{|0\rangle_{S_2} -|1\rangle_{S_2}}{\sqrt{2}} \otimes |\xi\rangle_{M_2},
    \label{p12}
\end{align}
\begin{align}
    |\psi^{+-}_7\rangle_{1,2,3,4} =& \frac{1}{\sqrt{2}} \Big( |0\rangle_{S_0} |0\rangle_{S_3} -  |1\rangle_{S_0} |1\rangle_{S_3} \Big) \otimes |\xi\rangle_{M_0} \otimes |\xi\rangle_{M_3}  \nonumber \\
    &\otimes \frac{|0\rangle_{S_1} + |1\rangle_{S_1}}{\sqrt{2}} \otimes |\xi\rangle_{M_1} \otimes \frac{|0\rangle_{S_2} -|1\rangle_{S_2}}{\sqrt{2}} \otimes |\xi\rangle_{M_2},
    \label{p22}
\end{align}
and
\begin{align}
     |\psi^{-+}_7\rangle_{1,2,3,4} =& \frac{1}{\sqrt{2}} \Big( |0\rangle_{S_0} |0\rangle_{S_3} -  |1\rangle_{S_0} |1\rangle_{S_3} \Big) \otimes |\xi\rangle_{M_0} \otimes |\xi\rangle_{M_3}  \nonumber \\
    &\otimes \frac{|0\rangle_{S_1} - |1\rangle_{S_1}}{\sqrt{2}} \otimes |\xi\rangle_{M_1} \otimes \frac{|0\rangle_{S_2} + |1\rangle_{S_2}}{\sqrt{2}} \otimes |\xi\rangle_{M_2}.
    \label{p22}
\end{align}

Now, the appropriate measurements are performed on the spins $S_0$ and $S_3$. For example, spin measurements corresponding the operators $A_0 = \sigma_x$ and $A_1 = -\sigma_z$  are performed on $S_0$, and spin measurements corresponding the operators $B_0 = (\sigma_x+\sigma_z)/\sqrt{2}$ and $B_1 = (\sigma_x-\sigma_z)/\sqrt{2}$ are performed on $S_3$. 
The statistics thus collected are used to test Bell nonlocality, in particular, the Bell-CHSH inequality. 

\section{Post-Selection and Bell Nonlocality:} \label{A2}
\paragraph{Local-realistic model and fair post-selection.}
Consider two distant observers, Alice and Bob, who share a hidden variable
\[
\lambda \in \Lambda \qquad \text{distributed according to } p(\lambda).
\]
On each run, Alice chooses an input (measurement setting) $x$ and obtains an outcome $a$; Bob chooses $y$ and obtains $b$.
A hidden-variable model is \emph{local} if for all $\lambda\in\Lambda$,
\begin{equation}\label{eq:local_factor}
p(a,b\mid x,y,\lambda)=p(a\mid x,\lambda)\,p(b\mid y,\lambda),
\end{equation}
equivalently
\[
p(a\mid x,y,\lambda)=p(a\mid x,\lambda),\qquad
p(b\mid x,y,\lambda)=p(b\mid y,\lambda).
\]
If, moreover, the model is \emph{deterministic} (reality), then for all $\lambda$ and all $x,y$,
\[
p(a,b\mid x,y,\lambda)\in\{0,1\}.
\]
The observable correlation is
\begin{equation}\label{eq:parent_corr}
p(a,b\mid x,y)=\sum_{\lambda\in\Lambda} p(\lambda)\,p(a,b\mid x,y,\lambda).
\end{equation}
By Bell's theorem, any correlation admitting the decomposition \eqref{eq:local_factor} satisfies the CHSH inequality and hence is Bell-local.

\vspace{0.5em}
Now suppose that, in addition to producing $(a,b)$ for inputs $(x,y)$, the parties also record outcomes
\[
m\in\mathcal{M},\qquad n\in\mathcal{N}
\]
of some \textbf{\emph{fixed} local measurements $M$ (Alice) and $N$ (Bob),} respectively (for concreteness one may take
$\mathcal{M}=\mathcal{N}=\{0,1\}$).
We consider the \emph{post-selected} correlation
\[
p(a,b\mid x,y,m,n),
\]
i.e., the statistics of $(a,b)$ conditioned on the event that the outcomes of $(M,N)$ are $(m,n)$.

\medskip
\paragraph{Fair post-selection (selection independence).}
We assume that the post-selection criterion is \emph{fair} in the following sense (for details see \cite{Brunner2014Bell}): conditioned on $\lambda$,
the probability of obtaining $(m,n)$ from the fixed measurements $(M,N)$ does not depend on the Bell-test inputs $(x,y)$, i.e.,
\begin{equation}
p(m,n\mid x,y,M,N,\lambda)=p(m,n\mid M,N,\lambda)\qquad \forall\,x,y,\lambda.\label{eq:fair_post1}
\end{equation}
Equivalently, the event ``$(m,n)$ occurs'' does not introduce any setting-dependent bias in the hidden variable distribution.
(If \eqref{eq:fair_post1} fails, then one enters the usual post-selection/detection loophole, where Bell violations can be
simulated by local models.) If the inputs $x,y$ are chosen in the causal future of the occurrence of $m,n$, and the choice of $x,y$ is made independently of $(m,n)$ and $\lambda$ (free choice, no retrocausality), then fair post-selection is guaranteed and we can write the above condition as:

\begin{equation}\label{eq:fair_post}
p(m,n\mid x,y,\lambda)=p(m,n\mid \lambda)\qquad \forall\,x,y,\lambda.
\end{equation}

\medskip
\paragraph{Subensembles induced by post-selection.}
Assume $p(m,n\mid M,N)>0$. Throughout, $M$ and $N$ are fixed local measurements producing outcomes $m$ and $n$, respectively. 
Hence we write $p(m,n)$ as shorthand for $p(m,n\mid M,N)$.
 Define the conditional (renormalized) hidden-variable distribution
\begin{equation}\label{eq:posterior_lambda}
p(\lambda\mid m,n)
=\frac{p(\lambda)\,p(m,n\mid \lambda)}{p(m,n)},
\end{equation}
where
\begin{equation}\label{eq:pmn_def}
p(m,n)=\sum_{\lambda\in\Lambda} p(\lambda)\,p(m,n\mid\lambda).
\end{equation}
In the deterministic case, $p(m,n\mid \lambda)\in\{0,1\}$ and one may equivalently introduce the subset
\begin{equation}\label{mn-model}
\Lambda_{mn}:=\{\lambda\in\Lambda:\ p(m,n\mid\lambda)=1\},
\end{equation}
so that \eqref{eq:posterior_lambda} reduces to $p(\lambda\mid m,n)
=\frac{p(\lambda)}{p(m,n)}$ a normalized distribution supported on $\Lambda_{mn}$.

\medskip
\paragraph{Post-selected correlations remain local.}
Using Bayes' rule we compute:
\begin{align}
p(a,b\mid x,y,m,n)
&=\sum_{\lambda\in\Lambda} p(\lambda\mid x,y,m,n)\,p(a,b\mid x,y,m,n,\lambda)\nonumber\\
&=\sum_{\lambda\in\Lambda} p(\lambda\mid m,n)\,p(a,b\mid x,y,m,n,\lambda). \label{eq:key_step}
\end{align}
Here the second line follows because: 

By Bayes' rule,
\begin{equation}\label{eq:bayes_dropMN}
p(\lambda\mid x,y,m,n)
=\frac{p(\lambda\mid x,y)\,p(m,n\mid x,y,\lambda)}{p(m,n\mid x,y)}.
\end{equation}
Assume \emph{free choice} (measurement independence),
\begin{equation}\label{eq:free_choice_dropMN}
p(\lambda\mid x,y)=p(\lambda),
\end{equation}
and \emph{fair post-selection} (no setting-dependent selection),
\begin{equation}\label{eq:fair_dropMN}
p(m,n\mid x,y,\lambda)=p(m,n\mid \lambda).
\end{equation}
Then
\begin{align}
p(m,n\mid x,y)
&=\sum_{\lambda} p(\lambda\mid x,y)\,p(m,n\mid x,y,\lambda)\nonumber\\
&=\sum_{\lambda} p(\lambda)\,p(m,n\mid \lambda)
=:p(m,n). \label{eq:pmn_dropMN}
\end{align}
Substituting \eqref{eq:free_choice_dropMN}, \eqref{eq:fair_dropMN}, and \eqref{eq:pmn_dropMN} into
\eqref{eq:bayes_dropMN}, we obtain
\begin{equation}\label{eq:key_posterior_dropMN}
p(\lambda\mid x,y,m,n)
=\frac{p(\lambda)\,p(m,n\mid \lambda)}{p(m,n)}
=:p(\lambda\mid m,n),
\end{equation}
which shows that conditioning on $(m,n)$ does not introduce any $x,y$-dependence in the posterior distribution of $\lambda$.

Now since from sub-hidden variable model as defined in Eq.(\ref{mn-model}) and using Eq.(\ref{eq:posterior_lambda}), we can safely write 

\begin{equation}
    p(\lambda|m,n) = 0 ~~\text{when}~~ \lambda \notin \Lambda_{mn}.
\end{equation}

From the above fact we can re-express Eq.(\ref{eq:key_step}) as

\begin{align}
p(a,b\mid x,y,m,n) &=\sum_{\lambda\in\Lambda_{mn}} p(\lambda\mid m,n)\,p(a,b\mid x,y,m,n,\lambda)\nonumber\\
&=\sum_{\lambda\in\Lambda_{mn}} p(\lambda\mid m,n)\,p(a,b\mid x,y,\lambda)\nonumber\\
&=\sum_{\lambda\in\Lambda_{mn}} p(\lambda\mid m,n)\,p(a\mid x,\lambda)\,p(b\mid y,\lambda) \label{eq:key_step(2)}
\end{align}
where, $p(a,b\mid x,y,m,n,\lambda)=p(a,b\mid x,y,\lambda)$ since $m,n$ is fixed and $p(a,b\mid x,y,\lambda)=p(a\mid x,\lambda)\,p(b\mid y,\lambda)$ follows from locality constraint as defined in Eq.(\ref{eq:local_factor}).

Equation \eqref{eq:key_step(2)} is precisely a local hidden-variable decomposition for the post-selected correlation.
Therefore, for every $(m,n)$ with $p(m,n\mid M,N)>0$, the correlation $p(a,b\mid x,y,m,n,M,N)$ is Bell-local. Hence, starting from a local (and in particular
local-deterministic) hidden-variable model, \emph{no fair post-selection on $(m,n)$ can generate a CHSH violation.}

\end{document}